\documentclass{emulateapj}

\usepackage{amsmath}
\usepackage{amssymb}

\slugcomment{Submitted to ApJL}

\shorttitle{Rotation curve of the Milky Way from Classical Cepheids}
\shortauthors{Mr\'oz et al.}

\begin{document}

\title{Rotation curve of the Milky Way from Classical Cepheids}

\author{Przemek~Mr\'oz$^{1}$\altaffilmark{$\dagger$}, Andrzej~Udalski$^{1}$, Dorota~M.~Skowron$^{1}$, Jan~Skowron$^{1}$, Igor~Soszy\'nski$^{1}$, Pawe\l{}~Pietrukowicz$^{1}$, Micha\l{}~K.~Szyma\'nski$^{1}$, Rados\l{}aw~Poleski$^{2,1}$, Szymon~Koz\l{}owski$^{1}$, and Krzysztof~Ulaczyk$^{3,1}$}

\affil{$^{1}$Warsaw University Observatory, Al. Ujazdowskie 4, 
00-478 Warszawa, Poland}
\affil{$^{2}$Department of Astronomy, Ohio State University, 140 W. 18th Ave., Columbus, OH 43210, USA}
\affil{$^{3}$Department of Physics, University of Warwick, Coventry CV4 7AL, UK}

\altaffiltext{$\dagger$}{Corresponding author: pmroz@astrouw.edu.pl}

\begin{abstract}
Flat rotation curves of spiral galaxies are considered as an evidence for dark matter, but the rotation curve of the Milky Way is difficult to measure. Various objects were used to track the rotation curve in the outer parts of the Galaxy, but most studies rely on incomplete kinematical information and inaccurate distances. Here, we use a sample of 773 Classical Cepheids with precise distances based on mid-infrared period-luminosity relations coupled with proper motions and radial velocities from \textit{Gaia} to construct the accurate rotation curve of the Milky Way up to the distance of $\sim 20$\,kpc from the Galactic center. We use a simple model of Galactic rotation to measure the rotation speed of the Sun $\Theta_0=233.6 \pm 2.8$\,km\,s$^{-1}$, assuming a prior on the distance to the Galactic center $R_0=8.122 \pm 0.031$\,kpc from the Gravity Collaboration. The rotation curve at Galactocentric distances $4\lesssim R\lesssim 20$\,kpc is nearly flat with a small gradient of $-1.34\pm 0.21$\,km\,s$^{-1}$\,kpc$^{-1}$. This is the most accurate Galactic rotation curve at distances $R>12$\,kpc constructed so far.
\end{abstract}


\section{Introduction}

Flat rotation curves of spiral galaxies provide evidence for dark matter \citep{rubin1980} or even ``new physics'' \citep{milgrom1983}, but the rotation curve of our Galaxy is notoriously difficult to measure, especially in the outer parts of the Milky Way. The most popular approach, the tangent-point method (e.g., \citealt{burton1978,clemens1985,fich1989,sofue2009,mcclure2016}), based on radio and mm observations of common molecules (H{\sc i} or CO), allows measuring the rotation curve within the solar orbit, although it is unreliable in the central regions of the Galaxy \citep{chemin2015}. The rotation curve outside the solar orbit can be measured with known distances and velocities of some tracers: H~{\sc ii} regions \citep{fich1989,brand1993}, Cepheids \citep{pont1994,pont1997,metzger1998,kawata2018}, open clusters \citep{hron1987}, or planetary nebulae \citep{durand1998}, but the current uncertainties are considerable (see Figure 1 of \citealt{sofue2009}), mostly because of poorly known distances. Such approach is prone to systematic errors, as usually only one component of the velocity vector (radial or tangential) is known and the circular rotation is assumed. As radial velocities and proper motions are measured relative to the Sun, both methods require independent information about the velocity of the Sun and distance to the Galactic center. See \citet{bhata2014}, \citet{pato2017}, \citet{russeil2017}, and references therein for recent data compilations.

A novel approach for constructing the Galactic rotation curve is presented by \citet{reid2009,reid2014} and \citet{honma2012}, who have measured accurate trigonometric parallaxes, proper motions and radial velocities of about 100 high-mass star-forming regions. They use the three-dimensional velocity information to calculate the rotation curve and to simultaneously estimate the velocity and location of the Sun. Their sample is relatively small and most of  objects they analyze are located in the northern part of the Galactic disk, which may introduce some bias. The local rotation curve was published by the \citet{katz2018}, who used the second \textit{Gaia} data release (\textit{Gaia}~DR2) to study motions of nearby stars, but their parallaxes are accurate in the solar neighborhood, within $2-3$\,kpc of the Sun.

Recently, \citet{udalski} presented the new OGLE Collection of Galactic Cepheids containing 1426 Classical Cepheids based on the survey of the Galactic plane carried out as part of the Optical Gravitational Lensing Experiment (OGLE). This data set more than doubled the number of known Galactic Cepheids. The survey covers over 2500 square degrees along the Galactic plane ($-170^{\circ}<l<+40^{\circ}$, $-6^{\circ}<b<+3^{\circ}$) and probes the Galactic disk out to its expected boundary ($\sim 20$\,kpc from the Galactic center). That sample, supplemented with previously known all-sky Cepheids, was used by \citet{skowron} to study the structure of the young Milky Way disk. 

Here, we complement distances to Cepheids from \citet{skowron} with the kinematical data (proper motions, radial velocities) to measure the three-dimensional velocities of Cepheids (Section \ref{sec:data}). We use a simple model of Galactic rotation to measure the velocity of the Sun (Section \ref{sec:model}) and to construct the accurate rotation curve of the Milky Way up to the Galactocentric distance of 20\,kpc (Section \ref{sec:rotation}).

\section{Data}
\label{sec:data}

\citet{skowron} measured accurate distances for 2177 Galactic Cepheids, using period-luminosity relations of \citet{wang2018} and mid-infrared light curves, which virtually removes the effects of interstellar extinction. We cross-matched Skowron et al.'s catalog with \textit{Gaia} DR2 \citep{gaia2016,gaia2018} and found that the full velocity information (proper motions and median radial velocities) is available for 832 objects\footnote{The data, as well as the modeling code, are publicly available at ftp://ftp.astrouw.edu.pl/ogle/ogle4/ROTATION\_CURVE/.}. 

We used distances measured by \citet{skowron}, as \textit{Gaia} parallaxes are not sufficiently accurate for many objects from our sample. Additionally, \citet{riess2018} and \citet{groene} showed that \textit{Gaia} parallaxes are systematically lower than accurate non-\textit{Gaia} parallaxes of Classical Cepheids by $-0.046 \pm 0.013$\,mas and $-0.049 \pm 0.018$\,mas, respectively. We also found a similar median offset of $-0.071 \pm 0.038$\,mas between \textit{Gaia} and \citeauthor{skowron}'s (\citeyear{skowron}) parallaxes (we compared distances of 251 Cepheids that have parallax uncertainties smaller than 10\%). The similar parallax zero-point offset, from $-0.029$ to $-0.082$\,mas, was found for other tracers (see \citealt{groene} and references therein). Typical distance uncertainties are of a few per cent. 

\begin{table*}
\centering
\caption{Best-fit model parameters}
\label{tab:params}
\begin{tabular}{|l|rrr|rrr|}
\hline \hline
Parameter & Model 1 & Model 2 & Model 3 & Model 1 & Model 2 & Model 3\\
& \multicolumn{3}{c|}{without prior on $R_0$} & \multicolumn{3}{c|}{with prior on $R_0$}\\
\hline 
$\Theta_0$ (km\,s$^{-1}$)                       & $221.3 \pm 3.6$ & $222.3 \pm 3.6$         & $221.5 \pm 3.6$   & $233.3 \pm 2.6$   & $233.6 \pm 2.6$   & $233.8 \pm 2.7$   \\
$\frac{d\Theta}{dR}$ (km\,s$^{-1}$\,kpc$^{-1}$) & $0.0$ (fixed)   & $-1.32 \pm 0.20$        & \dots             & $0.0$ (fixed)     & $-1.34 \pm 0.20$  & \dots \\
$R_0$ (kpc)                                     & $7.57 \pm 0.12$ & $7.60 \pm 0.11$         & $7.56 \pm 0.12$   & $8.09 \pm 0.03$   & $8.09 \pm 0.03$   & $8.09 \pm 0.03$   \\
$U_s$ (km\,s$^{-1}$)                            & $1.3 \pm 1.0$   & $1.6 \pm 1.0$           & $1.5 \pm 1.0$     & $1.4 \pm 1.0$     & $1.7 \pm 1.0$     & $1.7 \pm 1.0$     \\
$V_s$ (km\,s$^{-1}$)                            & $-5.2 \pm 2.2$  & $-3.4 \pm 2.3$          & $-2.2 \pm 2.3$    & $-6.2 \pm 2.2$    & $-4.4 \pm 2.2$    & $-3.4 \pm 2.3$    \\
$W_s$ (km\,s$^{-1}$)                            & $1.0 \pm 0.8$   & $1.0 \pm 0.8$           & $1.0 \pm 0.8$     & $1.0 \pm 0.8$     & $1.0 \pm 0.8$     & $1.0 \pm 0.8$     \\
$U_{\odot}$ (km\,s$^{-1}$)                      & $9.7 \pm 1.0$   & $10.0 \pm 1.0$          & $9.9 \pm 1.0$     & $9.8 \pm 1.0$     & $10.1 \pm 1.0$    & $10.1 \pm 1.0$    \\
$V_{\odot}$ (km\,s$^{-1}$)                      & $12.1 \pm 2.2$  & $12.1 \pm 2.1$          & $12.0 \pm 2.1$    & $12.2 \pm 2.1$    & $12.3 \pm 2.1$    & $12.1 \pm 2.2$    \\
$W_{\odot}$ (km\,s$^{-1}$)                      & $7.3 \pm 0.7$   & $7.3 \pm 0.7$           & $7.3 \pm 0.7$     & $7.3 \pm 0.7$     & $7.3 \pm 0.7$     & $7.3 \pm 0.7$     \\
$a_1$ (km\,s$^{-1}$)                            & \dots           & \dots                   & $222.8 \pm 3.6$   & \dots             & \dots             & $235.0 \pm 2.8$   \\
$a_2$                                           & \dots           & \dots                   & $0.88 \pm 0.05$   & \dots             & \dots             & $0.89 \pm 0.05$   \\
$a_3$                                           & \dots           & \dots                   & $1.31 \pm 0.06$   & \dots             & \dots             & $1.31 \pm 0.06$   \\
$\epsilon_l$ (km\,s$^{-1}$)                     & $13.5 \pm 0.4$  & $13.4 \pm 0.4$          & $13.3 \pm 0.4$    & $13.8 \pm 0.4$    & $13.7 \pm 0.4$    & $13.4 \pm 0.4$    \\
$\epsilon_b$ (km\,s$^{-1}$)                     & $7.6 \pm 0.2$   & $7.6 \pm 0.2$           & $7.6 \pm 0.2$     & $7.6 \pm 0.2$     & $7.6 \pm 0.2$     & $7.6 \pm 0.2$     \\
$\epsilon_r$ (km\,s$^{-1}$)                     & $14.6 \pm 0.5$  & $14.3 \pm 0.5$          & $14.3 \pm 0.5$    & $14.6 \pm 0.5$    & $14.2 \pm 0.5$    & $14.2 \pm 0.5$    \\
\hline
$\Omega_0=\Theta_0/R_0$ (km\,s$^{-1}$kpc$^{-1}$)    & $29.22 \pm 0.33$ & $29.25 \pm 0.33$ & $29.28 \pm 0.33$ & $28.84 \pm 0.31$ & $28.88 \pm 0.31$ & $28.92 \pm 0.32$ \\
$(\Theta_0+V_{\odot})/R_0$ (km\,s$^{-1}$kpc$^{-1}$) & $30.82 \pm 0.20$ & $30.84 \pm 0.19$ & $30.88 \pm 0.19$ & $30.35 \pm 0.16$ & $30.40 \pm 0.16$ & $30.41 \pm 0.16$ \\
$\Delta\ln\mathcal{L}$                              & 0.0 & 21.3 & 35.3 & 0.0 & 21.4 & 34.2\\
$\rho_{R_0,\Theta_0}$                               & 0.74             & 0.73             & 0.73             & 0.28             & 0.27             & 0.26 \\
\hline
\end{tabular}

\raggedright 
\textbf{Note.} Model~1: flat rotation curve $\Theta(R)=\Theta_0=\mathrm{const}$, model~2: linear rotation curve $\Theta(R)=\Theta_0+\frac{d\Theta}{dR}(R-R_0)$, model~3: the universal rotation curve \citep{persic1996}, see equations (\ref{eq:per1})--(\ref{eq:per2}). $\Delta\ln\mathcal{L}$ is the log-likelihood improvement relative to the model~1. $\rho_{R_0,\Theta_0}$ is the correlation coefficient between $R_0$ and $\Theta_0$. We used the following Gaussian priors on the motion of the Sun: $U_{\odot}=11.1 \pm 1.3$\,km\,s$^{-1}$, $V_{\odot}=12.2 \pm 2.1$\,km\,s$^{-1}$, and $W_{\odot} = 7.3 \pm 0.7$\,km\,s$^{-1}$ \citep{schonrich2010}. We consider models with and without the Gaussian prior on $R_0=8.112 \pm 0.031$\,kpc \citep{gravity2018}.
\end{table*}

Known Cepheids located in binary systems\footnote{http://www.konkoly.hu/CEP/intro.html} \citep{szabados2003} were not included in the modeling. For the final models, we also removed a few objects with residual velocities at least $4\sigma$ larger than the mean, where $\sigma$ is the dispersion of residuals. These can be unrecognized binary Cepheids (with wrong \textit{Gaia} astrometric solution) or variables of other type that were mistaken with Cepheids. We were left with 773 objects. The radial velocities of Cepheids show variations with amplitudes up to 30\,km\,s$^{-1}$ with the pulsation period \citep{joy1937,stibbs1955}. Radial velocities reported in the \textit{Gaia}~DR2 are median values of single-transit measurements. Cepheids from our sample were observed from 2 to 44 times with the median number of seven visits. Small number of single observations is usually reflected by large error bars, although in some cases, the uncertainties may be underestimated (if the measurements happened to be collected near the same pulsation phase). Thus, for the modeling, we added in quadrature a constant value (about $14\,$km\,s$^{-1}$) to the reported radial velocity uncertainties.

\section{Modeling}
\label{sec:model}

We use a simple model of circular rotation of the Milky Way. For each Cepheid, with known Galactic coordinates (longitude $l$ and latitude $b$) and heliocentric distance $D$, we calculate the expected radial and tangential velocities and compare them with observations. 

Our model has the following free parameters: $R_0$ -- distance of the Sun to the Galactic center, $(U_s,V_s,W_s)$ -- mean noncircular motion of the source in a Cartesian Galactocentric frame, $(U_{\odot},V_{\odot},W_{\odot})$ -- solar motion with respect to the local standard of rest (LSR), and one to three parameters that describe the shape of the rotation curve. We follow the notation from Appendix of \citet{reid2009}: $U_i$ is the velocity component toward the Galactic center, $V_i$ -- along the Galactic rotation, $W_i$ -- toward the North Galactic pole. We consider three analytical rotation curves: $\Theta(R)=\Theta_0=\mathrm{const}$ (model~1) and $\Theta(R)=\Theta_0+\frac{d\Theta}{dR}(R-R_0)$ (model~2), where $\Theta_0$ and $\frac{d\Theta}{dR}$ are parameters and $R$ is the distance to the Galactic center. The third model is based on a universal rotation curve introduced by \citet{persic1996}, which describes in a simple way contributions from the stellar disk and dark matter halo to the total rotation velocity. That model has three parameters: $a_1$ -- the rotation speed at the optical radius $R_{\rm opt}$ of the galaxy, $a_2 = R_{\rm opt} / R_0$, and $a_3$ -- the ``velocity core radius'' (in units of $R_{\rm opt}$), $a_2$ and $a_3$ define the shape of the rotation curve:
\begin{align}
\label{eq:per1}
\Theta^2_d(x) &= a_1^2 b \frac{1.97x^{1.22}}{(x^2+0.78^2)^{1.43}}\\
\Theta^2_h(x) &= a_1^2 (1-b) x^2 \frac{1+a_3^2}{x^2+a_3^2}\\
\Theta(x) &= \sqrt{\Theta^2_d(x)+\Theta^2_h(x)}\\
\label{eq:per2}
x &= R / R_{\rm opt} = (R / R_0) / a_2.
\end{align}
We adopt $b=0.72$ \citep{persic1996,reid2016}.

The total velocity of the Cepheid is $(U_s,V_s+\Theta(R),W_s)$. Let $\beta$ be the angle between the Sun and the source as viewed from the Galactic center (see Figure 9 of \citealt{reid2009}). We rotate the velocity vector through the angle $-\beta$ and subtract the velocity of the Sun:
\begin{align}
\label{eq:eq1}
U_1 &= U_s \cos\beta + (V_s + \Theta(R))\sin\beta - U_{\odot},\\
V_1 &= -U_s\sin\beta + (V_s + \Theta(R))\cos\beta - V_{\odot} - \Theta_0, \\
W_1 &= W_s - W_0.
\end{align}
The radial velocity $V_r$ and tangential velocities in Galactic coordinates ($V_l$ and $V_b$) can be calculated as follows:
\begin{align}
V_l &= V_1 \cos{l} - U_1 \sin{l}, \\
V_b &= W_1 \cos{b} - (U_1\cos{l}+V_1\sin{l}) \sin{b}, \\
V_r &= W_1 \sin{b} + (U_1\cos{l}+V_1\sin{l}) \cos{b} .
\label{eq:eq2}
\end{align}
We maximize the following likelihood function:
\begin{equation}
\ln\mathcal{L}=-\frac{1}{2}\sum\limits_{i={1\dots N}}\sum\limits_{j={l,b,r}}\left(\frac{(V_{i,j}-V_{i,j,\mathrm{model}})^2}{\sigma^2_{V_{i,j}}+\epsilon_j^2}+\ln(\sigma^2_{V_{i,j}}+\epsilon_j^2)\right),
\label{eq:lik}
\end{equation}
where $\sigma_{V_{i,j}}$ is the velocity uncertainty and $\epsilon_l$, $\epsilon_b$, and $\epsilon_r$ are additional parameters that describe the scatter in $V_l$, $V_b$, and $V_r$ (the scatter of residuals is much larger than the original error bars owing to the peculiar (noncircular) motion of stars).

The best-fit parameters are found by maximizing the likelihood function using the simplex approach \citep{nelder1965}. 
The uncertainties are estimated using the Markov chain Monte Carlo (MCMC) technique \citep{foreman2013} and represent 68\% confidence range of marginalized posterior distributions, see Table \ref{tab:params}. As we found that the velocity of the Sun with respect to the LSR is poorly constrained by the data, we used the following Gaussian priors: $U_{\odot}=11.1 \pm 1.3$\,km\,s$^{-1}$, $V_{\odot}=12.2 \pm 2.1$\,km\,s$^{-1}$, and $W_{\odot} = 7.3 \pm 0.7$\,km\,s$^{-1}$ \citep{schonrich2010}. We assumed uniform priors on other parameters.

We found that $\Theta_0$ and $R_0$ are strongly correlated. Their correlation coefficient, calculated using MCMC chains, is equal to about 0.73 (Table~\ref{tab:params}). Our best estimates of $R_0\approx 7.6$\,kpc are smaller than the most accurate current measurement from the Gravity Collaboration \citep[$R_0=8.122 \pm 0.031$\,kpc;][]{gravity2018}. Other recent determinations also favor the larger value: $R_0=7.93 \pm 0.14$\,kpc \citep{chu2018} and $R_0=8.20 \pm 0.09$\,kpc \citep{mcmillan2017}. To understand the reason of this difference, we added a Gaussian prior on $R_0$. We found that models with the prior on $R_0$ are disfavored by $2\Delta\ln\mathcal{L}=19.4$, but most this signal can be attributed to two stars EX~Mus ($D=17.5\pm0.8$\,kpc, $A_K=0.08$\,mag) and V800~Aql ($D=18.5\pm1.0$\,kpc, $A_K=0.24$\,mag), both of which are located far from the Sun and their distance may be affected by systematic errors (mostly the interstellar extinction $A_K$ in the $K$ band). If these two stars are removed from the sample, models with $R_0=8.122$\,kpc are disfavored by only $2\Delta\ln\mathcal{L}=8.4$. As there is a priori no reason to remove these two stars, we prefer to use models with priors on $R_0$.

Moreover, models without a constraint on $R_0$ produce the angular speed of the Sun about the Galactic center ($(\Theta_0+V_{\odot})/R_0$, see Table~\ref{tab:params}) that is in tension with the accurate measurement of the proper motion of Sgr~A* \citep[$30.24 \pm 0.12$\,km\,s$^{-1}$\,kpc;][]{reid2004}. Using the $R_0$ measured by the Gravity Collaboration removes this tension.
Then, the estimated $\Theta_0$ raises to $233.6 \pm 2.6$\,km\,s$^{-1}$ (model~2) and $233.8 \pm 2.7$\,km\,s$^{-1}$ (model~3), see Table~\ref{tab:params}.

\begin{figure}
\includegraphics[width=0.5\textwidth]{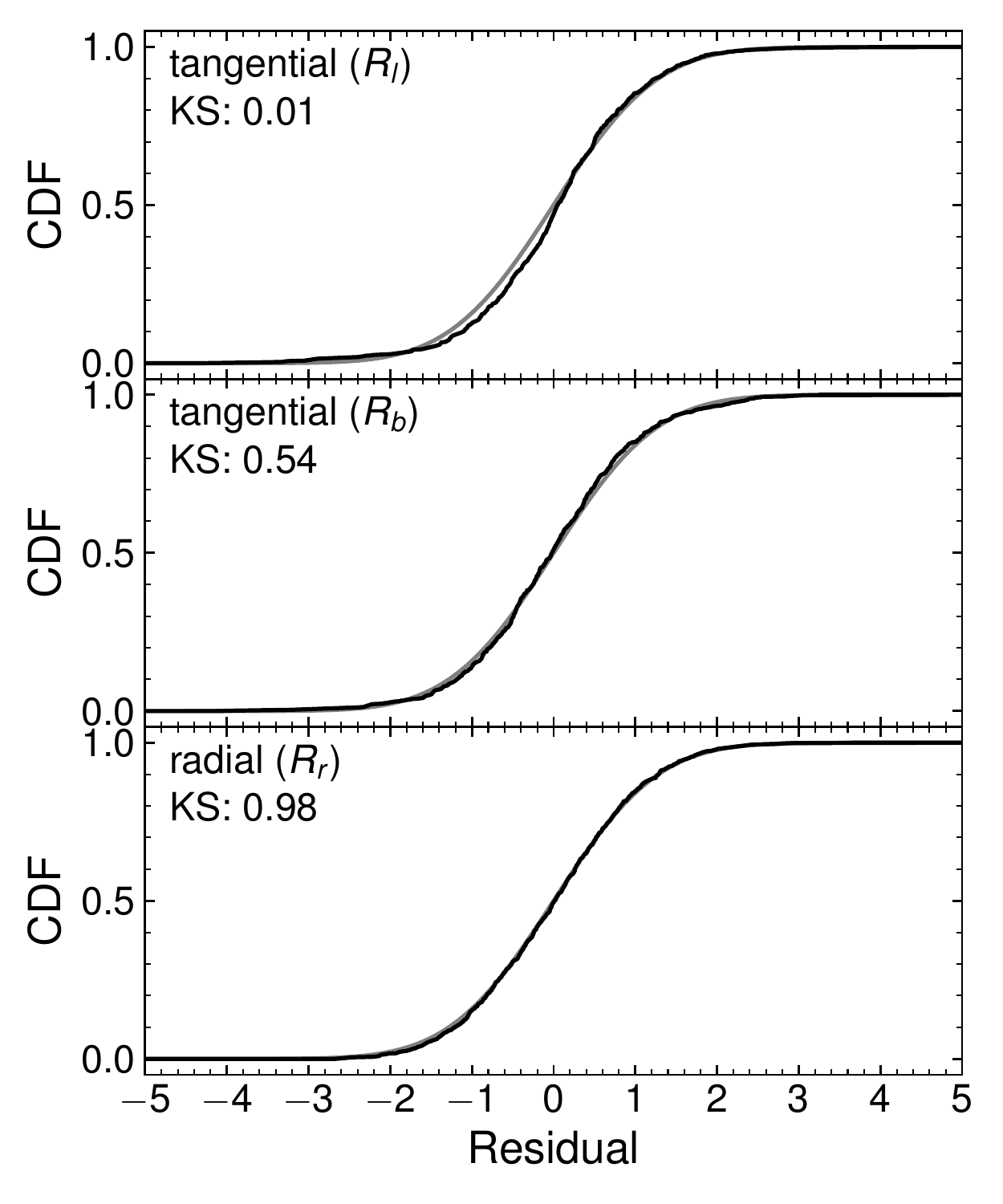}
\caption{The black line shows the cumulative distribution of residuals from the best-fit model (model~2 with the prior on $R_0$). This distribution is compared to a standard Gaussian distribution, we quote the $p$-value of the Kolmogorov-Smirnov (KS) test. The gray line shows the cumulative distribution function of the normal distribution.}
\label{fig:cdf}
\end{figure}

The cumulative distribution of residuals $R_{i,j}=(V_{i,j}-V_{i,j,\mathrm{model}})/\sqrt{\sigma^2_{V_{i,j}}+\epsilon_j^2}$ from the best-fit model (model~2 with the prior on $R_0$) is shown in Fig.~\ref{fig:cdf}. The residuals (of all three velocity components) follow the Gaussian distribution well.

\begin{figure*}
\includegraphics[width=\textwidth]{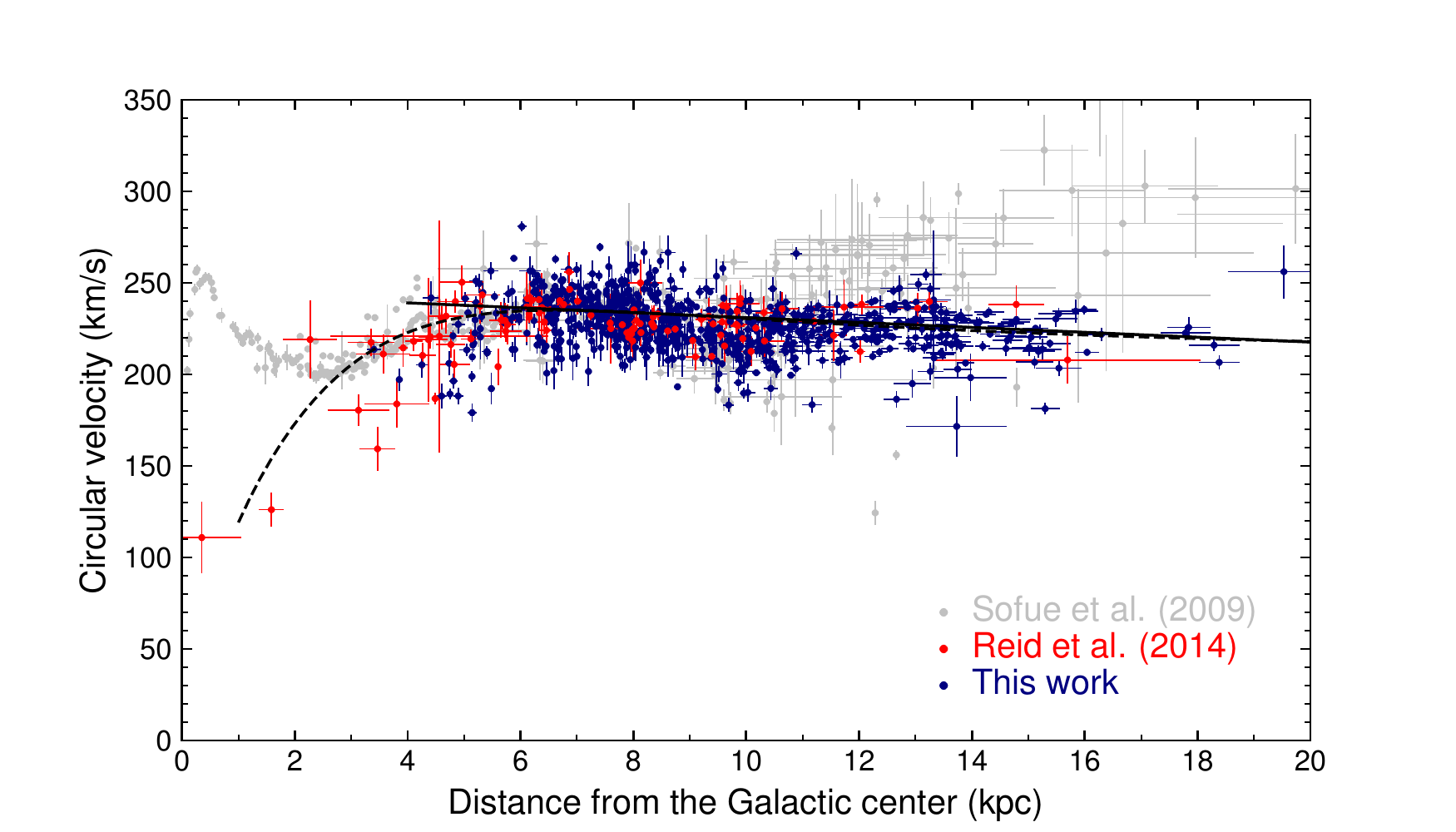}
\caption{Rotation curve of the Milky Way for Cepheids assuming $R_0=8.09$\,kpc and $\Theta_0=233.6$\,km\,s$^{-1}$ (model~2). Red data points represent high mass star forming regions \citep{reid2014}. Grey data points are taken from \citet{sofue2009} and were rescaled to new $(R_0,\Theta_0)$ using formula $V_{\rm new}=V_{\rm old}+\frac{R}{8.0}\left(\Theta_0-200\right)$. Solid and dashed lines show the best-fitting models (linear and universal, respectively).} 
\label{fig:rot}
\end{figure*}

\begin{figure}
\includegraphics[width=0.5\textwidth]{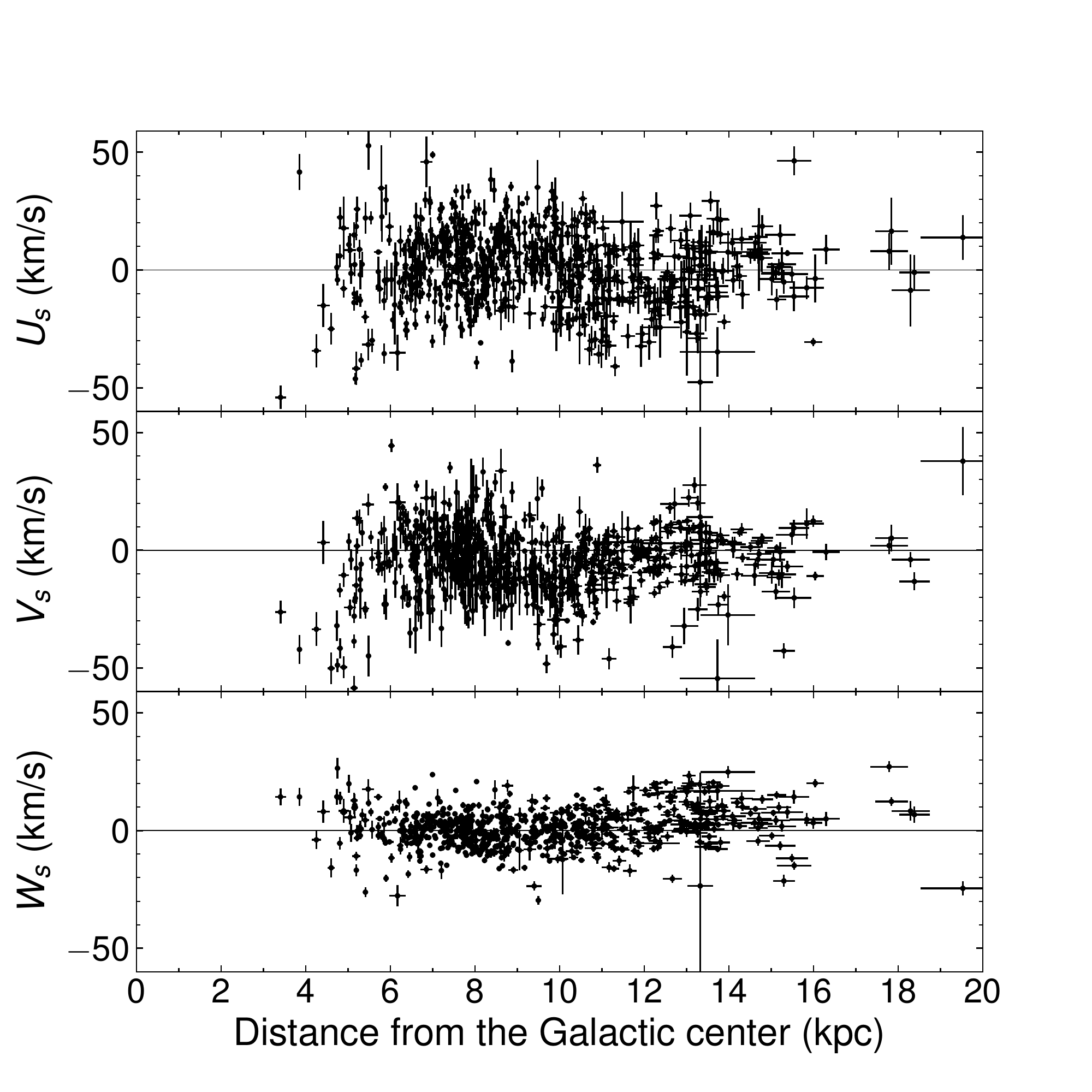}
\caption{Peculiar (noncircular) motions of Cepheids, after subtracting the model of Galactic rotation (model 2, linear rotation curve). $U_s$ is the velocity component toward the Galactic center, $V_s$ -- along the Galactic rotation, and $W_s$ toward the North Galactic pole.} 
\label{fig:resid}
\end{figure}

We found that both linear and universal rotation curves describe that data much better than a simple constant rotation curve, with log-likelihood improvement $2\Delta\ln\mathcal{L}$ of 42.8 and 68.4, respectively. Linear and universal rotation curve models have similar maximal log-likelihoods, but the latter model is slightly preferred. This preference is mainly caused by a few Cepheids with velocities lower than 200\,km\,s$^{-1}$, which are located at distances $\approx 4$\,kpc from the Galactic center (Figure~\ref{fig:rot}). Similarly, \citet{reid2014} found that velocities of masers closest to the Galactic center ($R\lesssim 5$\,kpc) deviate from the flat rotation curve. The universal rotation curve model of \citet{persic1996}, however, does not agree with observations collected by \citet{sofue2009} for $R\lesssim3$\,kpc (Figure~\ref{fig:rot}). These measurements, obtained with the tangent-point method may be unreliable as argued by \citet{chemin2015}. The current sample of Cepheids is too small to credibly distinguish between the two models.

To assess how distance uncertainties influence the final parameters, we carried out Monte Carlo simulations. For each Cepheid, we drew a new distance from the normal distribution and repeated our modeling procedure. We conducted 100 trials, in which we found the additional spread of $\Theta_0$ and $d\Theta/dR$ of 0.8\,km\,s$^{-1}$ and 0.05\,km\,s$^{-1}$\,kpc$^{-1}$, respectively. We add these quantities in quadrature to the uncertainties from Table~\ref{tab:params}, finding $\Theta_0=233.6\pm2.8$\,km\,s$^{-1}$ and $d\Theta/dR=-1.34\pm0.21$\,km\,s$^{-1}$\,kpc$^{-1}$ for model~2.

Residuals from the best-fit models are shown in Figure~\ref{fig:resid}, separately for radial, azimuthal, and vertical velocity components. Error bars of many individual objects are much lower than the scatter ($\sigma_U = 16$\,km\,s$^{-1}$, $\sigma_V=14$\,km\,s$^{-1}$, $\sigma_W=8$\,km\,s$^{-1}$), likely because of peculiar (noncircular) motion of stars. Some Cepheids may be unrecognized members of binary systems.

The measured rotation speed of the Sun $\Theta_0$ is in good agreement with previous determinations. \citet{reid2014} found $\Theta_0 = 240 \pm 8$\,km\,s$^{-1}$ and $R_0 = 8.34 \pm 0.16$\,kpc based on parallaxes and proper motions of high-mass star-forming regions. We measured a slightly smaller velocity of the Sun, but the angular rotation of the Sun about the Galactic center $(\Theta_0+V_{\odot})/R_0=30.40 \pm 0.16$\,km\,s$^{-1}$\,kpc is similar to that found by \citet{reid2014} ($30.57\pm0.43$\,km\,s$^{-1}$\,kpc). \citet{reid2004} measured the proper motion of Sagittarius A* of $30.24 \pm 0.12$\,km\,s$^{-1}$\,kpc, which corresponds to $\Theta_0+V_{\odot}= 241.9 \pm 1.0$\,km\,s$^{-1}$ for $R_0 = 8$\,kpc. The angular velocity of circular rotation of the Sun ($\Omega_0 = \Theta_0 / R_0 =28.88 \pm 0.31$\,km\,s$^{-1}$\,kpc$^{-1}$) in our model is consistent with \textit{Hipparcos} ($27.19 \pm 0.87$\,km\,s$^{-1}$\,kpc$^{-1}$; \citealt{feast1997}) and \textit{Gaia} ($27.2 \pm 0.6$\,km\,s$^{-1}$\,kpc$^{-1}$; \citealt{bovy2017}) measurements. 

\section{Galactic rotation curve}
\label{sec:rotation}

We use parameters ($R_0$,$\Theta_0$) from Table \ref{tab:params}, model 2 to construct the rotation curve of the Milky Way. We convert radial and tangential heliocentric velocities to the Galactocentric velocity by adding the motion of the Sun (equations (\ref{eq:eq1}--\ref{eq:eq2})). The resulting rotation curve is shown in Figure~\ref{fig:rot}, where we also plotted earlier data from \citet{sofue2009}, rescaled to new ($R_0$,$\Theta_0$). Both data sets agree well up to a distance of 10--11\,kpc from the Galactic center. Previous studies (e.g., \citealt{sofue2009,russeil2017}) found that the rotation curve outside 12\,kpc is nearly constant or even rising (although its precise shape may depend on the choice of $R_0$ and $\Theta_0$), but these data were affected by large uncertainties and small number of observations (Figure~\ref{fig:rot}). Our rotation curve is nearly flat with a small gradient of $-1.34 \pm 0.21$ km\,s$^{-1}$\,kpc$^{-1}$, contrary to some earlier claims that the rotation of Cepheids is Keplerian \citep{gnacinski2018}. 

Classical Cepheids are excellent tracers of the rotation curve in the outer parts of the Milky Way disk. Our rotation curve outside 12\,kpc is more accurate than in any previous studies \citep{sofue2009,reid2014} and can be used to constrain the distribution of dark matter in the Milky Way. Currently, our sample includes only 128 Cepheids at Galactocentric distances greater than 12\,kpc (out of nearly 600 Cepheids with $R>12$\,kpc from \citealt{skowron}). Future \textit{Gaia} data releases, as well as a dedicated spectroscopic survey of Cepheids, can provide more accurate insight into rotation of the outer parts of the Milky Way disk.

\section*{Acknowledgments}

We thank the anonymous referee for constructive comments which helped us to improve the paper.
P.M. acknowledges support from the Foundation for Polish Science (Program START). The OGLE project has received funding from the National Science Centre, Poland, grant MAESTRO 2014/14/A/ST9/00121 to A.U. I.S. is also supported by the Polish National Science Centre grant MAESTRO 2016/22/A/ST9/00009.
This work has made use of data from the European Space Agency (ESA) mission
{\it Gaia} (\url{https://www.cosmos.esa.int/gaia}), processed by the {\it Gaia}
Data Processing and Analysis Consortium (DPAC,
\url{https://www.cosmos.esa.int/web/gaia/dpac/consortium}). Funding for the DPAC
has been provided by national institutions, in particular the institutions
participating in the {\it Gaia} Multilateral Agreement.

\bibliographystyle{aasjournal}
%\bibliography{pap}

\end{document}